\documentclass[11pt,a4paper]{article}

\usepackage{amsmath,amsthm,amssymb,amsfonts}

\usepackage{authblk}
\usepackage{setspace}
\usepackage{mathrsfs}
\usepackage{cases}
\usepackage{graphicx}

\numberwithin{equation}{section}
\newcommand{\ii}{{\rm i}}
\newcommand{\ee}{{\rm e}}

\begin{document}

\title{Can a particle detector cross a Cauchy horizon?\footnote{Based on a talk given by the author at ``VII Black Holes Workshop" at the University of Aveiro, Portugal, 18 - 19 December 2014.}}
\author{Benito A. Ju\'arez-Aubry\thanks{Work done in collaboration with Jorma Louko.}}
\affil{School of Mathematical Sciences\\ University of Nottingham\\ Nottingham NG7 2RD\\ UK \\
{\tt pmxbaju@nottingham.ac.uk}}
\date{February 2015, Revised April 2015}

\maketitle


\begin{abstract}
%
%
Cauchy horizons are well known to exhibit instabilities in classical
spacetime dynamics and singularities in quantum field theory. We
analyse the response of an Unruh-DeWitt particle detector that falls
towards a Cauchy horizon, in terms of the specifics of the horizon,
the choice of the quantum state and the specifics of the detector's
trajectory. As a prototype, we study in detail the case for the $1+1$
Reissner-Nordstr\"om black hole with a scalar field in the Hartle-Hawking
state. Comparisons are made with the response of a detector that
falls into a Schwarzschild-like singularity.
\end{abstract}

\singlespacing

\section{Introduction}
\label{sec:intro}

Causality is one of the main pillars of physics. At a classical level, it is directly related to the well-posedness of the dynamical equations cast as an initial value problem in which the evolution of the solution is determined fully by the initial conditions specified on a spacelike hypersurface. At a quantum level, causality is best understood at the level of the algebra of observables. Given two elements in the algebra of observables with spacelike-separated support, then the commutator between the two elements must vanish.

Intimately related to the concept of causality is that of global hyperbolicity. Indeed, every globally hyperbolic spacetime $(M,g)$ is strongly causal (and in fact stably causal). This means that whenever there exists a Cauchy hypersurface $\Sigma \subset M$, no causal curve can come arbitrarily close to intersecting itself. The global hyperbolicity of the spacetime is a necessary and sufficient condition for the dynamical equations of classical fields to be well-posed as an initial value problem on the whole manifold $M$ with initial data specified as configuration of fields and field velocities or momenta (Lagrangian and Hamiltonian formalisms resp.) on a Cauchy hypersurface of the spacetime.\footnote{The reader may review standard references \cite{Wald:1984rg,Hawking:1973uf} for  technical details.}

Having said this, it may be surprising that some of the most important solutions in General Relativity are not globally hyperbolic. Notably, most members of the Kerr-Newman family are among these solutions. Such spacetimes contain Cauchy horizons as boundaries of the maximal Cauchy development of a given achronal hypersurface, after which the causal structure of the spacetime is violated.  

On this line, a great deal of effort has been spent in understanding the nature of the Cauchy horizons in General Relativity. Initial observations lead to the \textit{strong cosmic censorship conjecture} \cite{Simpson:1973ua}, which claims that for generic initial data the spacetime is inextendible beyond the maximal Cauchy development. The classical analysis of the experience of an observer crossing a Cauchy horizon was first studied in detail by Chandrasekhar and Hartle \cite{Chandrasekhar:1982}. On assessing the strong cosmic censorship conjecture, most of the efforts concentrated in the Reissner-Nordstr\"om black hole, where spherical symmetry is available. In this case, given polynomially decaying data for a dynamical Maxwell-Einstein-scalar system settling to a Reissner-Nordstr\"om black hole, the spacetime is extendible with a $C^0$-metric, falsifying the conjecture. However, not all the physical invariants are finite and, in particular, the Hawking mass goes to infinity. This is the mass inflation scenario \cite{Poisson:1989zz, Poisson:1990eh, Dafermos:2002ka, Dafermos:2003wr}. See ref. \cite{Costa:2014zha} for a recent analysis including a non-zero cosmological constant.

Thus far, all of the above is concerned with classical aspects of the problem of causality. The problem that we would like to address is that of the effect of the presence of quantum fields on non-globally hyperbolic spacetimes. After all, matter, like the fundamental interactions and (very plausibly) gravity, is quantum. The particular question that we ask is: what is the experience of an observer approaching a Cauchy horizon as it interacts with matter? We take the standpoint that a quantum scalar field lives on a background spacetime which solves the Einstein field equations and that the backreaction is weak, such that the natural language for the problem is that of quantum field theory in curved spacetimes.\footnote{See standard literature \cite{Birrell:1982ix, Wald:1995yp} for an introduction.}

This is a valid approximation and the success of this approach is documented in the literature, including the discovery of black hole Hawking radiation \cite{Hawking:1974sw, Gibbons:1977mu}, cosmological particle creation \cite{Parker:1969au, Parker:1971pt, Zeldovich:1971mw} and the Unruh effect \cite{Unruh:1976db, DeWitt:1979, Unruh:1983ms, Crispino:2007eb, Bisognano:1976za}. 

An observer following their worldline in the spacetime has access to the power spectrum of the vacuum noise of the field by measuring the transition probability (response function) of an ensemble of particle detectors \cite{Takagi:1986kn}. The goal of this paper is to analyse the behaviour of the transition probability per unit time as a particle detector approaches a Cauchy horizon. We will work in $1+1$ dimensions and with a massless scalar field in the Hartle-Hawking-Israel vacuum, where we can use conformal invariance to obtain analytical expressions. Our analysis will concentrate in Reissner-Nordstr\"om-like solutions.

The agenda is the following: In Section \ref{sec:detector} we will develop the technology of the Unruh-DeWitt detector in $1+1$ dimensions. Section \ref{sec:blackhole} will be concerned with the computation of the transition rate in the vicinity of the Cauchy horizon of a Reissner-Nordstr\"om-like solution. Finally, our concluding remarks are found in Section \ref{sec:conclusions}.

\section{Particle detectors in $1+1$ dimensions}
\label{sec:detector}

The Unruh-DeWitt model that we would like to describe consists of a physical system with Hamiltonian $H = H_\phi + H_D + H_I$, where $H_\phi$ is the free Hamiltonian associated to the scalar field in $1+1$ dimensions, $H_D$ is the Hamiltonian of a point-like detector and $H_I$ is the interaction Hamiltonian between the detector and the field. The internal space of the detector is a two-dimensional Hilbert space with energy eigenstates labelled by $\lbrace | \downarrow \rangle, | \uparrow \rangle \rbrace$, such that $H_D | \downarrow \rangle = 0$ and $H_D | \uparrow \rangle = \omega | \uparrow \rangle$. The Unruh-DeWitt detector in $n+1$ dimensions, $n \geq 1$, has an interaction Hamiltonian given by $H_\text{int} = c \mu(\tau) \chi(\tau) \phi(\mathsf{x}(\tau))$, where $c$ is a small time-independent coupling constant compared to the scales of the problem, $\mu$ is usually thought of as the monopole moment of the detector and $\chi$ is a smooth switching function of compact support.

We wish to work in $1+1$ dimensions in order to exploit the conformal symmetry of the theory of massless fields. However, the coupling that we have just written does not lead to well-defined expressions. For example, when computing the transition probability of the detector \cite{Birrell:1982ix} from the state $| \downarrow \rangle$ to the state $| \uparrow \rangle$ coupled to a massless field found initially in an arbitrary Hadamard state one obtains an infinite ambiguity that needs to be subtracted \textit{ad hoc} to make sense of the result of the computation. We can avoid this situation in two ways. By working exclusively with massive fields, in which case the conformal symmetry is destroyed, or by modifying the interaction Hamiltonian coupling. In a previous article \cite{Juarez-Aubry:2014jba} we argue that the latter strategy is more natural. We will work with the derivative-coupling detector with interaction 

\begin{equation}
H_\text{int} = c \mu(\tau) \chi(\tau) \dot{\phi}(\mathsf{x}(\tau)),
\end{equation}
\\
where $\tau$ is the proper time of the detector along its worldline and the overhead dot indicates a derivative with respect to $\tau$. The expression for the transition probability of the detector is now well defined for any massive or massless field in any Hadamard state. The transition probability can be computed to linear order in the coupling $c$. To this order, the transition probability is equal to the response function of the detector multiplied by a function that depends only on the internal details of the detector. The response function is given by

\begin{equation}
\mathcal{F}(\omega) \doteq \int_{-\infty}^\infty \! d\tau' \, \int_{-\infty}^\infty \! d\tau'' \, \ee^{-\ii \omega (\tau'-\tau'')} \chi(\tau') \chi(\tau'') \partial_{\tau'} \partial_{\tau''} \mathcal{W}(\tau',\tau''),
\label{F1}
\end{equation}
\\
where $\mathcal{W}(\tau',\tau'') \doteq \langle 0 | \phi(\mathsf{x}(\tau')),\phi(\mathsf{x}(\tau''))|0 \rangle$ is the pullback of the Wightman function of the Klein-Gordon field along the detector worldline. The derivatives on the right hand side of eq. \eqref{F1} are understood as distributional derivatives. This expression is finite as $\chi$ is smooth and compactly supported. This expression can be given in terms of integrals of integrable functions\footnote{Under certain circumstances additional divergences may arise from boundary conditions in the problem \cite{Juarez-Aubry:2014jba}. In this case, the expression in eq. \eqref{F2} will continue to be of distributional nature, but we will not face this problem in this article.} using the short-distance behaviour of the Hadamard two-point function \cite{Decanini:2005eg}. The expression reads \cite{Juarez-Aubry:2014jba}

\begin{align}
\mathcal{F}(\omega) & = -\frac{\omega}{2} \int_{-\infty}^\infty \! du \left[ \chi(u) \right]^2 + \frac{1}{\pi}\int_{0}^\infty \! \frac{ds}{s^2}  \int_{-\infty}^\infty \! du \chi(u) \left[\chi(u) - \chi(u-s) \right] \nonumber \\
& + 2 \int_{-\infty}^\infty \! du \, \int_{0}^\infty \! ds \, \chi(u) \chi(u-s) \text{Re} \left[\ee^{-\ii \omega s} \partial_{u} \partial_{u-s} \mathcal{W}(u,u-s) + \frac{1}{2 \pi s^2} \right].
\label{F2}
\end{align}

In practice, it is simpler to work in the sharp-switching limit by deforming the switching function $\chi(u)$ into $\chi_0(u) = \Theta(u-\tau_0)\Theta(\tau-u)$ as we let the switching time go to zero \cite{Louko:2007mu}. Here $\Theta$ is the Heaviside function. By doing this, a logarithmic constant divergence is encountered for the transition probability and it is more convenient to work with the transition rate, i.e. the transition probability per unit time. In this limit, the transition rate is

\begin{align}
\dot{\mathcal{F}}(\omega,\tau, \tau_0) & = -\omega \Theta(-\omega) + \frac{1}{\pi} \left[\frac{\cos(\omega\Delta \tau)}{\Delta \tau} + |\omega|\text{si}(|\omega|\Delta \tau) \right] \nonumber \\
& + 2 \int_{0}^{\Delta\tau} \! ds \, \text{Re} \left[\ee^{-\ii \omega s} \partial_{\tau} \partial_{\tau-s} \mathcal{W}(\tau,\tau-s) + \frac{1}{2 \pi s^2} \right],
\label{Fdot}
\end{align}
\\
where we have spelled out the explicit dependence of the transition rate in the measurement time $\tau$ and the switch-on time $\tau_0$ on the left hand side. Here $\text{si}$ is the sine integral \cite{NIST} and $\Delta \tau \doteq \tau - \tau_0$ is the total interaction proper time.

With eq. \eqref{Fdot} at hand, we can compute the transition rate of a detector approaching a Cauchy horizon and performing a measurement at proper time $\tau$ in its vicinity. This allows us to infer what an observer following the same worldline would feel as it interacts with the quantum matter $\phi$ sitting on the spacetime.

\section{Crossing the Reissner-Nordstr\"om Cauchy horizon}
\label{sec:blackhole}

\subsection{The $1+1$ Reissner-Nordstr\"om black hole}

We will start by reviewing the Reissner-Nordstr\"om spacetime. This will serve also the purpose of introducing the relevant notation. The $3+1$ Reissner-Nordstr\"om spacetime is a spherically symmetric, asymptotically flat solution satisfies the Einstein-Maxwell equation. It depends on two parameters associated with the total mass $M$ and total charge $Q$ of the spacetime. Whenever $Q^2 \leq M^2$ it is a black hole. In our analysis, we exclude the extremal scenario and deal with the $Q^2 < M^2$ black hole. Spherical symmetry allows one to obtain a $1+1$ solution starting out from the standard $3+1$ spacetime by dropping the angular components of the solution. The $1+1$ metric in the exterior region is

\begin{equation}
g = -F(r) dt^2 + F(r)^{-1} dr^2,
\label{extRN}
\end{equation}
\\
with

\begin{align}
F(r)  = \frac{(r-r_+)(r-r_-)}{r^2}, \hspace{2cm} r_\pm  = M \pm (M^2-Q^2)^{1/2}.
\label{Fr}
\end{align}

From eq. \eqref{extRN} it is clear that the exterior region is static with $\xi = \partial_t$ being a Killing vector field. From eq. \eqref{Fr} we can see that the solution contains two horizons, at $r = r_+$ and $r = r_-$ and that a curvature timelike singularity is located at $r = 0$.

It is well known that the Reissner-Nordstr\"om black hole can be extended in a tower of copies of itself by introducing a set of Kruskal-Szekeres coordinates like so: Starting from the exterior region $r > r_+$, define the null coordinates $(u, v) \doteq (\tau-r^*, \tau+r^*)$ with $dr^* = dr/F(r)$ the tortoise coordinate. The Kruskal coordinates for $r > r_-$ are defined by $(U,V) \doteq (-\exp(-\kappa_+ u), \exp(\kappa_+ v))$, with $\kappa_\pm = F'(r_\pm)/2$ the surface gravities at $r_\pm$. The prime denotes differentiation with respect to the argument. The coordinates $(U,V)$ can then be extended to four regions as indicated in Table \ref{exterior}.

\begin{table}[h]
\begin{center}
\begin{tabular}{ l l c r }
  & Region & sgn$(U)$ & sgn$(V)$ \\
  \hline
  I: Exterior &$r_+ < r$ & $-$ & + \\
  II: Black hole &$r_-< r < r_+$ & + & + \\
  I': Isometric exterior &$r_+ < r$ & + & $-$ \\
  II': White hole &$r_-< r < r_+$ & $-$ & $-$ 
\end{tabular}
\caption{Coordinate chart in the region $r_-<r<\infty$.}
\label{exterior}
\end{center}
\end{table}

\begin{figure}[ht]
\begin{center}
\includegraphics[height=8cm]{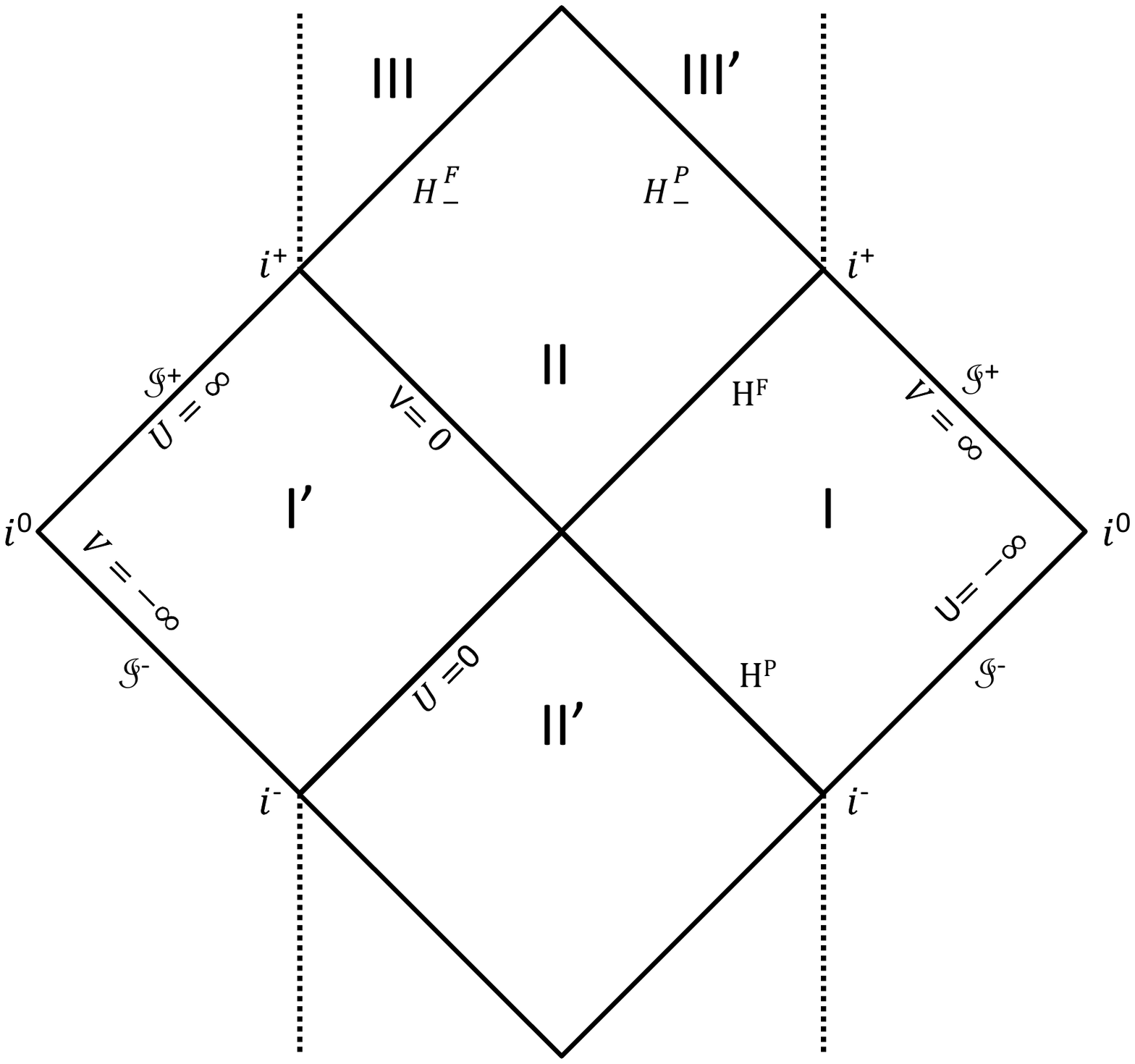}
\caption{Part of the conformal diagram of the $1+1$ Reissner-Nordstr\"om non-extremal black hole. The bifurcate Killing horizon at $r_+$ is $H^\text{P} \cup H^\text{F}$. The bifurcate Killing horizon at $r_-$ is $H^\text{P}_- \cup H^\text{F}_-$ and part of it is displayed in the future of the diagram. The singularities in regions III and III' are indicated by dotted lines. The diagram extends to the past and future.}
\label{RNConformal}
\end{center}
\end{figure}

There is a bifurcate Killing horizon at $r_+$, $H^\text{P} \cup H^\text{F}$, where $H^\text{P}$ is generated by $\partial_U$ and $H^\text{F}$ by $\partial_V$. To cover the region $0 \leq r<r_-$ one can introduce the coordinates in region II $(\tilde{u},\tilde{v})\doteq(\tilde{r}^*-\tilde{t},\tilde{r}^*+\tilde{t})$ and define the Kruskal-Szekeres coordinates for $r_-<r<r_+$ by $(U_-,V_-) =( -\exp(\kappa_- \tilde{u}), -\exp(\kappa_- \tilde{v}))$. This allows one to construct a coordinate chart for $0 <r<r_+$ in a similar way. The bifurcate Killing horizon at $r_-$, $H_-^\text{P} \cup H_-^\text{F}$, is generated by $\partial_{U_-}$ and $\partial_{V_-}$. There are left and right timelike curvature singularities located at $U_-V_- = -1$. See figure \ref{RNConformal} for details.

We now proceed to perform the scalar field quantisation in this spacetime.

\subsection{Scalar field quantisation}

The Hartle-Hawking-Israel (HHI) state in the Reissner-Nordstr\"om spacetime is defined to be regular in the region $\text{I} \cup H^\text{P} \cup \text{II} \cup \text{I'} \cup H^\text{F} \cup \text{II'}$. This differs drastically from the picture in the Schwarzschild black hole, where the state is regular in the whole Kruskal manifold. A second HHI state can be obtained using the bifurcate Killing horizon at $r_-$ which is then regular in $\text{II} \cup H_-^\text{P} \cup \text{III} \cup \text{II''} \cup H_-^\text{F} \cup \text{III'}$ if appropriate boundary conditions are set at $U_- V_- = -1$. For an observer coming from the exterior and falling into the black hole region, it is the first HHI state the one that is relevant. Our analysis will deal with this vacuum state.

The standard Hartle-Hawking-Israel state for the theory of a massless field can be obtained in $1+1$ dimensions in closed form by exploiting the conformal symmetry of the theory, modulo an infrared ambiguity that does not play any role in our calculations. In the case of Reissner-Nordstr\"om, by writing the metric in the region $r>r_-$ as $g = F(r)\left( \kappa_+^2 U V \right)^{-1} dU dV$, we can identify the region $\text{I} \cup H^\text{P} \cup \text{II} \cup \text{I'} \cup H^\text{F} \cup \text{II'}$ with the Minkowski flat spacetime using the conformal factor $\Omega^2 = -F(r)\left(\kappa_+^2 UV \right)^{-1}$. In this way, the Klein-Gordon equation simplifies as

\begin{equation}
\Box \phi = 0 \Leftrightarrow \partial_U \partial_V \phi = 0
\end{equation}
\\
and the canonical quantisation procedure is identical to that of Minkowski spacetime. The vacuum state, like all other quasi-free states, can be given in terms of the two-point function exclusively. It is given in Kruskal-Szekeres coordinates by

\begin{equation}
\langle 0 | \phi(\mathsf{x}) \phi(\mathsf{x'}) | 0 \rangle = \mathcal{W}(\mathsf{x},\mathsf{x'}) = -\frac{1}{4 \pi} \ln \left[(\epsilon+i \Delta U) (\epsilon + i \Delta V) \right] + \text{div}.
\end{equation}

By examining eq. \eqref{F2} and \eqref{Fdot} one can notice that the infrared divergent term, which we denote div, does not play any role in our computation, since it is a constant term and it vanishes when taking the derivatives in the right hand side of the transition rate formula [see eq. \eqref{Fdot}]. This is the virtue of the derivative-coupling Unruh-DeWitt detector that we advocate for $1+1$ spacetime dimensions \cite{Juarez-Aubry:2014jba}.

We now proceed to analyse the transition rate of a geodesically infalling observer.

\subsection{Freely-falling detector transition rate}

We would like to stress that, while Reissner-Nordstr\"om is our physical motivation, nowhere in our calculations do we need to specify the detailed form of $F(r)$ in the metric tensor \cite{Juarez-Aubry:2015}. It suffices to say that it is a smooth function with the following properties:

\begin{subequations}
  \begin{align}
    F(r)& =
    \begin{cases}
      > 0, & \text{if}\ r>r_+, \\
      < 0, & \text{if}\ r_-<r<r_+, \\
      > 0, & \text{if}\ r<r_-.
    \end{cases}
    \\
    F'(r) & = 
    \begin{cases}
      > 0, & \text{if}\ r=r_+, \\
      < 0, & \text{if}\ r=r_-.
    \end{cases}
  \end{align}
\end{subequations}

Because $F \in C^\infty(\mathbb{R})$, $F(r_\pm) = 0$, as it is the case in Reissner-Nordstr\"om. 

We let a geodesic observer fall from region I into region III and we compute its transition rate as it approaches the Cauchy horizon at $r_-$ on the left. In the exterior region, the geodesics are given by the integral curves of

\begin{subequations}
\begin{align}
\dot{t} &= \frac{E}{F(r)}, \label{t-eq} \\
\dot{r} &= -\sqrt{E^2-F(r)}, \label{r-eq}
\end{align}
\label{eqmotion} 
\end{subequations}
\\
where the dot stands for a derivative with respect to the proper time of the detector. The transition probability per unit (proper) time of the detector at proper time $\tau$ when the detector has been switched on in the exterior region at time $\tau_0$ is given by eq. \eqref{Fdot}. Integrating by parts we obtain

\begin{align}
\dot{\mathcal{F}}_\text{H}(\omega, \tau, \tau_0) & = - 2 \cos\left(\omega \Delta \tau\right)\partial_\tau \mathcal{W}_\text{H}(\tau,\tau_0)\nonumber \\
& + \lim_{\tau' \rightarrow \tau }2 \cos\left(\omega(\tau-\tau')\right) \left[\partial_\tau \mathcal{W}_\text{H}(\tau,\tau')+\frac{1}{2 \pi (\tau-\tau')} \right] \nonumber \\
& - 2 \int_{\tau_0}^\tau \! d \tau' \, \omega \sin \left(\omega(\tau-\tau')\right) \left[\partial_\tau \mathcal{W}_\text{H}(\tau,\tau')+\frac{1}{2 \pi (\tau-\tau')} \right] + \mathcal{O}(1),
\label{hhilim}
\end{align}
\\
where the order $\mathcal{O}(1)$ is with respect to the quantity $(\tau_h - \tau)$, which is small as the detector approaches the Cauchy horizon at proper time $\tau_h$. We shall show that the transition rate diverges proportionally to $(\tau_h-\tau)^{-1}$, preventing an observer to access the region $r < r_-$. The boundary contributions of the integral can by studied by means of an asymptotic expansion, by which one finds that

\begin{align}
\dot{\mathcal{F}}_\text{H}(\omega, \tau, \tau_0) & =  -\left[ \frac{1 +(\cos(\omega \Delta \tau)-1) \kappa_+/\kappa_-}{4 \pi } + \mathcal{O}\left((\tau_h-\tau)^{|\kappa_+/\kappa_-|} \right) \right]\left(\tau_h - \tau \right)^{-1} \nonumber \\
& - 2 \int_{\tau_0}^\tau \! d \tau' \, \omega \sin \left(\omega(\tau-\tau')\right) \left[\partial_\tau \mathcal{W}_\text{H}(\tau,\tau')+\frac{1}{2 \pi (\tau-\tau')} \right].
\label{hhiboundary}
\end{align}

Obtaining the contribution coming from the integral is more involved. The second term in eq. \eqref{hhiboundary} can be written as

\begin{align}
\frac{\omega}{2 \pi} \frac{\dot{U}(\tau)}{U(\tau)} \int_{\tau_0}^{\tau_h} \! d \tau' \,  \frac{\sin \left(\omega(\tau-\tau')\right)}{1 - U(\tau')/U(\tau)} + \mathcal{O}(1).
\end{align}

The key point is that the integral in the expression above is of order $\mathcal{O}(1)$. The main argument \cite{Juarez-Aubry:2015} accounts to finding an integrable function that bounds the integrand and applying the dominated convergence theorem in the limit $\tau \rightarrow \tau_h$. Using this technique the integral can be solved explicitly and we obtain that

\begin{align}
\dot{\mathcal{F}}_\text{H}(\omega,\tau,\tau_0) & = -\frac{1}{4 \pi} \left[ 1 +  \frac{\kappa_+}{\kappa_-}  + \textit{o}(1)\right] \frac{1}{\tau_h - \tau}.
\label{FdotFinal}
\end{align}

It is clear that the transition rate diverges as an inverse power of the small quantity $\tau_h - \tau$ along the geodesically infalling trajectory. The leading divergence is independent of the initial incoming velocity of the observer, since the parameter $E$ from the geodesic equations \eqref{eqmotion} does not play any role to leading order. This is a feature shared with a detector's response as it falls into the singularity of a Schwarzschild black hole in $1+1$ dimensions. We stress that this result is independent of the details of $F(r)$ and that the only input needed are the conditions \eqref{Fr}, which englobe a larger family of spacetimes than the Reissner-Nordstr\"om non-extremal black hole. 

\section{Concluding remarks}
\label{sec:conclusions}

By calculating the transition rate of a particle detector approaching the Cauchy horizon of the Reissner-Nordstr\"om black hole as it interacts with a massless scalar field, we have shown that the region $r < r_-$ is inaccessible for an observer in the limit of no back-reaction even when the interaction with the field is very weak. In terms of the proper time of the detector, the response per unit time diverges proportionally to $(\tau_h -\tau)^{-1}$ as $\tau \rightarrow \tau_h$. To leading order, this divergent behaviour is comparable to that of a detector approaching the spacelike $1+1$ Schwarzschild singularity, as in this case the leading divergence is equal to $1/[6\pi(\tau_\text{sing} - \tau)]$ as $\tau \rightarrow \tau_\text{sing}$ for a geodesically infalling observer \cite{Juarez-Aubry:2014jba}.  Inspection of eq. \eqref{FdotFinal} suggests that the behaviour is different whenever $\kappa_+ = - \kappa_-$. It can be shown \cite{Juarez-Aubry:2015} that for these critical values the transition rate is finite.

While our results have been obtained in $1+1$ dimensions, we have reasons to believe that they shed light to the $3+1$ dimensional experience, although in this case we expect the transition rate divergence to be weaker. To support this argument, we can compare the results in $3+1$ dimensions of a static detector crossing the Cauchy horizon of the Rindler spacetime embedded in the Minkowski spacetime right wedge as it interacts with a scalar field in the Rindler vacuum. For an inertial Unruh-DeWitt detector following the trajectory $x^\mu = (\tau, \tau_h,0,0)$ in $3+1$ dimensions crossing the future branch of the Rindler spacetime at proper time $\tau_h$, it is known that the transition rate diverges like $\ln(1-\tau/\tau_h)$ as the future Rindler horizon is approached \cite{Louko:2007mu}. A similar computation can be performed in $1+1$ dimensions with the derivative-coupling detector. We found that the divergence for the trajectory $x^\mu = (\tau, \tau_h)$ in $1+1$ dimensions is, instead, proportional to $(\tau_h-\tau)^{-1}$ with subleading logarithmic divergences \cite{Juarez-Aubry:2015}. The explanation comes from taking derivatives of the $1+1$ Wightman function. Although the short distance singularity structure of the twice-differentiated $1+1$ Wightman function is of the same order as its $(3+1)$-dimensional counterpart, it seems that the derivatives make the former less regular as the boundary of the domain of regularity of the Green function is approached.

A calculation that also confirms the conclusions that an observer cannot cross a Cauchy horizon comes from the computation of the vacuum expectation value of the energy momentum tensor in the HHI state. It is possible to estimate the energy that a local observer feels as it interacts with the energy-momentum of the scalar field using conformal techniques \cite{Davies:1976ei, Wald:1978ce}. In this way, one can compute the energy along the worldline of an observer,

\begin{equation}
\mathcal{E} \doteq \int \! d \tau' \, \chi(\tau') \dot{\gamma}^a(\tau') \dot{\gamma}^b(\tau')  \langle \mathcal{T}_{ab}(\tau') \rangle_{\text{ren}},
\end{equation}
\\
a quantity that is useful, for example, in the study of the quantum energy inequalities \cite{Fewster:1999gj}. The rate of divergence of $\mathcal{E}$ can be estimated explicitly as the observer approaches the Cauchy horizon for smooth and sharp test functions $\chi$ with compact support  \cite{Juarez-Aubry:2015}. It may also be of interest to compute the transition probability using a smooth switching function, using recent techniques for smoothly switched detectors \cite{Fewster:2015}, and compare these results with the sharp switching case.

As a final word, all of our estimates can be applied for a field in the Unruh vacuum state, defined regions I and II and across the event horizon, for the observer falling into region III. This follows from the decoupling of left and right moving modes in $1+1$ dimensions.


\section*{Acknowledgments}

The author thanks Jorma Louko, who is a collaborator in this work, for his comments on a draft version of this contribution. Thanks are also due to the organisers of the ``VII Black Holes Workshop'' for their hospitality. The author is supported by Consejo Nacional de Ciencia y Tecnolog\'ia (CONACYT), Mexico, REF 216072/311506.

\end{document}